\documentclass[aps,prb,onecolumn,superscriptaddress,showpacs,notitlepage]{revtex4-1}

\usepackage{graphicx}
\usepackage{amsmath}
\usepackage[section]{placeins}
\usepackage{subfig}
\usepackage{epstopdf}

\begin{document}


\title{ Inertial amplification of continuous structures: \\
Large band gaps from small masses}


\author{Niels M. M. Frandsen}
\email[]{nimmfr@mek.dtu.dk}
\affiliation{Department of Mechanical Engineering, The Technical University of Denmark, Denmark}

\author{Osama R. Bilal}
\altaffiliation{Currently at: Department of Mechanical and Process Engineering, ETH Zürich, Switzerland}
\affiliation{Department of Aerospace Engineering Sciences, University of Colorado Boulder, Boulder, CO 80309, USA}

\author{Jakob S. Jensen}
\affiliation{Department of Electrical Engineering, The Technical University of Denmark, Denmark}

\author{Mahmoud I. Hussein}
\affiliation{Department of Aerospace Engineering Sciences, University of Colorado Boulder, Boulder, CO 80309, USA}


\date{\today}

\begin{abstract}
Wave motion in a continuous elastic rod with a periodically attached inertial-amplification mechanism is investigated. The mechanism has properties similar to an ``inerter'' typically used in vehicle suspensions, however here it is constructed and utilized in a manner that alters the intrinsic properties of a continuous structure.
The elastodynamic band structure of the hybrid rod-mechanism structure yields band gaps that are exceedingly wide and deep when compared to what can be obtained using standard local resonators, while still being low in frequency.
With this concept, a large band gap may be realized with as much as twenty times less added mass compared to what is needed in a standard local resonator configuration.
The emerging inertially enhanced continuous structure also exhibits unique qualitative features in its dispersion curves. These include the existence of a characteristic double-peak in the attenuation constant profile within gaps and the possibility of coalescence of two neighbouring gaps creating a large contiguous gap.
\end{abstract}


\maketitle

\section{Introduction}
The band structure of a material represents the relation between wavenumber (or wave vector) and frequency, and thus it relates the spatial and temporal characteristics of wave motion in the material. This relation is of paramount importance in numerous disciplines of science and engineering such as electronics, photonics and phononics. \cite{Kittel1976}
It is well-known that periodic materials exhibit gaps in the band structure, \cite{Brillouin1953} referred to as \emph{band gaps} or \emph{stop bands}. In these gaps, waves are attenuated whereby propagating waves are effectively forbidden. Their defining properties are the frequency range, i.e., position and width, as well as the depth in the imaginary part of the wavenumber spectrum, which describes the level of attenuation. 

In the realm of elastic wave propagation, band gaps are usually created by two different phenomena: Bragg scattering or local resonance. 
Bragg scattering occurs due to the periodicity of a material or structure, where waves scattered at the interfaces cause coherent destructive interference, effectively cancelling the propagating waves. Research on waves in periodic structures dates, at least, back to Newton's attempt to derive a formula for the speed of sound in air, see e.g., Chapter 1 in Ref. \onlinecite{Brillouin1953} for a historical review before the 1950's. Later review papers on wave propagation in periodic media include Refs. \onlinecite{Elachi1976,Mead1996,Hussein2014}.

The concept of local resonance is based on the transfer of vibrational energy to a resonator, i.e., a part of the material/structure that vibrates at characteristic frequencies. Within structural dynamics, the concept dates back, at least, to Frahm's patent application \cite{Frahm1911} and since, dynamic vibration absorbers and tuned mass dampers have been areas of extensive research within structural vibration suppression. 
In the field of elastic band gaps, the concept of local resonance is often considered within the framework of periodic structures, as presented in the seminal paper of Liu et al., \cite{Liu2000} where band-gaps are created for acoustic waves using periodically distributed internal resonators. 
The periodic distribution of the resonators does not change the local resonance effects, however it does introduce additional Bragg scattering at higher frequencies, as well as allow for a unit-cell wave based description of the medium.
Local resonance has also been used in the context of attaching resonators to a continuous structure, such as a rod \cite{Yu2006a}, beam \cite{Yu2006} or a plate \cite{Pennec2008,Wu2008} in order to attenuate waves by creating band gaps in the low frequency range. A problem with this approach in general, which has severely limited proliferation to industrial applications, is that the resonators need to be rather heavy for a practically significant gap to open up.

Another means for creating band gaps is by the concept of inertial amplification (IA) as proposed by Yilmaz and collaborators in Refs. \onlinecite{Yilmaz2007,Acar2013}. In this approach, which has received less attention in the literature, inertial forces are enhanced between two points in a structure consisting of a periodically repeated mechanism. 
This generates \emph{anti-resonance} frequencies, where the enhanced inertia effectively cancels the elastic force; see e.g. Ref. \onlinecite{Yilmaz2006} where two levered mass-spring systems are analysed for their performance in generating stop bands.
While it is possible to enhance the inertia between two points by means of masses, springs and levers, a specific mechanical element, \emph{the inerter}, \cite{Smith2002} was created as the ideal inertial equivalent of springs and dampers, providing a force proportional to the relative acceleration between two points. 
This concept, while primarily used in vehicle suspension systems, \cite{Chen2009} has been utilized in Refs. \onlinecite{Yilmaz2007,Acar2013} in the context of generating band gaps in lattice materials by inertial amplification where the same underlying physical phenomenon is used for generating the anti-resonance frequencies. 
The frequency responses to various harmonic loadings were obtained, numerically and experimentally, and low-frequency, wide and deep band gaps were indeed observed for these novel lattice structures. In Ref. \onlinecite{Yuksel2015}, size and shape optimization is shown to increase the band gap width further, as illustrated by a frequency-domain investigation.

Until now, both inerters and inertial amplification mechanisms have been used as a backbone structural component in discrete or continuous systems. 
In this paper, we propose to use inertial amplification to generate band gaps in conventional continuous structures, by attaching light-weight mechanisms to a host structure, such as a rod, beam, plate or membrane, without disrupting its continuous nature (therefore not obstructing its main structural integrity and functionality). With this approach, we envision the inertial amplification effect to be potentially realized in the form a \emph{surface coating}, to be used for sound and vibration control.

For proof of concept, we consider a simple one-dimensional case by analyzing an elastic rod, with an inertial amplification mechanism periodically attached.
The mechanism is inspired by that analyzed in Ref. \onlinecite{Yilmaz2007}, however the application to a continuous structure increases the practicality and richness of the problem considerably and several novel effects are illustrated. 

Our investigation focuses mainly on the unit-cell band-structure characteristics. However, we also compare our findings from the analysis of the material problem to transmissibility results for structures comprising a finite number of unit cells. 
The finite systems are modelled by the finite-element (FE) method.

\section{Model\label{sec:model}}
In order to utilize the concept of inertial amplification in a surface setting as proposed, the mechanisms should be much smaller than the host structure, such that their distributed attachment does not change the main function of the structure, nor occupy a significant amount of space. Fulfilling this constraint requires a relatively large effect with only a modest increase in mass.

Considering the ideal mechanical element, the \emph{inerter}, we know that the factor of proportionality, the inertance, can be much larger than the actual mass increase, as demonstrated experimentally in Refs. \onlinecite{Smith2002,Papageorgiou2009}.
We propose to utilize the same effect using a mechanism similar to the one considered in Refs. \onlinecite{Yilmaz2007,Acar2013}. Our two-dimensional interpretation of the system may be viewed as a plate with a comparably small inertial amplification mechanisms distributed over the host-structure. In principle, the distributed effect of the mechanisms, in the long wave-limit, reduces to the notion of an inertially-modified constitutive relation in the elastodynamic equations.


\subsection{Model reduction}
\begin{figure} 
\centering
\includegraphics[width=0.43\textwidth]{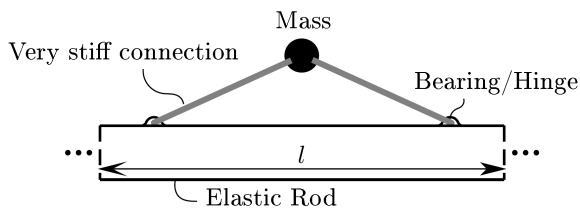}
\caption{1D version of the proposed concept.\label{fig:1D}}
\end{figure}

\begin{figure} 
\centering
\includegraphics[width=0.4\textwidth]{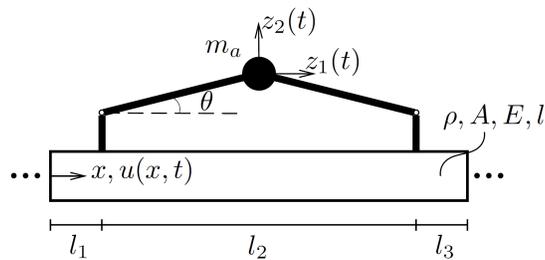}
\caption{Hybrid continuous-discrete model.\label{fig:hybrid}}
\end{figure}

In this study, we restrict ourselves to a one-dimensional structure with an inertial amplification mechanism attached, as illustrated in Fig. \ref{fig:1D}, where the mechanism is attached to the rod with bearings. A similar bearing is used at the top connection, such that, ideally, no moment is transferred through the mechanism.
This ensures that the connecting links do not deform, but move the amplification mass by rigid-body motion. 
The 1D-system is simplified further to a hybrid model consisting of a continuous, elastic bar  and a discrete mechanism as seen in Fig. \ref{fig:hybrid}, as this allows for a rigorous analytical formulation for the underlying dynamics.

The bar has Young's modulus $E$, cross-sectional area $A$, mass density $\rho$ and unit-cell length $l = l_1+l_2+l_3$, while the amplification mass is denoted $m_a$ and $\theta$ is the amplification angle. 
In Fig. 3, heavy lines indicate rigid connections and the corners between vertical and inclined rigid connections are moment-free hinges, hence the motion of the amplification mass  quantified by $z_1(t)$ and $z_2(t)$, is governed by the motion at the attachment points $u(x_1,t)$, $u(x_2,t)$ and the amplification angle $\theta$, where $x_1 = l_1$ and $x_2 = l_1 + l_2$.
It is noted that the model in Fig. \ref{fig:hybrid} will be unaffected by the mechanism in the static limit, $\omega = 0$. This is in agreement with the desire to produce a force that is proportional to the relative acceleration, thus changing the effective inertia of the system. From a physical standpoint, any increase in static stiffness of the mechanism would arise from frictional stiffness in the bearings or at the top point, however it is outside the scope of this work to include these residual stiffness effects, among other things, since they are assumed to be small.

\begin{figure} 
\centering
\includegraphics[width=0.4\textwidth]{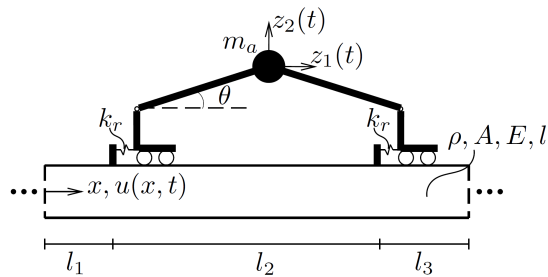}
\caption{Hybrid model including connection flexibility.\label{fig:localsystem}}
\end{figure}

The inertial amplification model in Fig. \ref{fig:hybrid} assumes rigid connections between the rod and the mechanism. Should the connections be flexible as illustrated in Fig. \ref{fig:localsystem}, the mechanism can work as both a local resonator as well as an inertial amplifier, depending on the specific parameters of the system. The local-resonance (LR) system in Fig. \ref{fig:localsystem} recovers the ideal inertial amplification system of Fig. \ref{fig:hybrid} in the limit $k_r \rightarrow \infty$. 
The inertial amplification system in Fig. \ref{fig:hybrid} is the main system investigated in this paper, while the system including the connection flexibility is used for two purposes: to illustrate that inertial amplification can occur even when connections are non-ideal, i.e., flexible, and to compare the behaviour of the inertial amplification mechanism to an equivalent, local-resonator-type system.

All the analytical formulations in this paper are based on the differential equation of a rod,
\begin{equation}
\rho \ddot{u}(x,t) = \sigma'(x,t) + f(x,t)
\label{eq:bardl}
\end{equation}
where $u(x,t)$ is the longitudinal displacement, $\sigma(x,t)$ is the normal stress while $\ddot{()} = \partial^2()/\partial t^2$ and $()' = \partial()/\partial x$. 
The body force $f(x,t)$ will not be present in the material problem formulation considering infinite domains.
The rod is considered to be homogeneous, however a layered rod would pose no additional difficulty in terms of the transfer matrix method described in Section \ref{ssec:transfermatrix}, since the method is applicable to layered materials.\cite{Mochan1987,Esquivel-Sirvent1994,Hussein2006} The rod is further assumed to be linear elastic with infinitesimal strains, which provides the constitutive relation
\begin{equation}
\sigma(x,t) = E\varepsilon(x,t) = Eu'(x,t)
\label{eq:constitutive}
\end{equation}
where $\varepsilon(x,t)$ is the longitudinal strain in the rod.

\subsection{Mechanism equations\label{ssec:mechanism}}
Before considering the hybrid systems in Figs. \ref{fig:hybrid} and \ref{fig:localsystem}, the mechanisms are considered with constraint forces applied to account for the rod. These constraint forces are determined in terms of the mechanism parameters, whereby the effect of the mechanism on the rod is given in terms of these constraint forces.

\subsubsection{Kinematics}
\begin{figure} 
\centering
\includegraphics[width=0.42\textwidth]{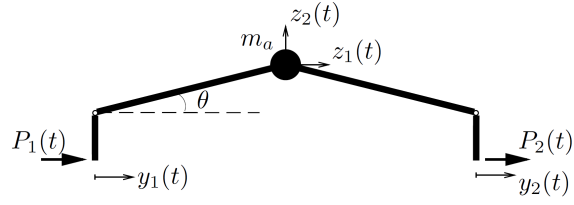}
\caption{IA mechanism with constraint forces.\label{fig:mechanismIA}}
\end{figure}

Considering the isolated inertial amplification mechanism in Fig. \ref{fig:mechanismIA}, the motions $z_1$ and $z_2$ can be determined in terms of $y_1$, $y_2$ and $\theta$.
In the appendix, the full non-linear kinematic relations are derived. In this paper we consider the linearized version, whereby $z_1$ and $z_2$ are determined as
\begin{subequations}
\begin{align}
z_1 &= \frac{1}{2}(y_2+y_1) \\
z_2 &= \frac{1}{2}\cot\theta(y_2-y_1)
\end{align}
\end{subequations}
as seen in Ref. \onlinecite{Yilmaz2007} as well. 

\subsubsection{Constraint forces} 
\begin{figure} 
\centering
\includegraphics[width=0.42\textwidth]{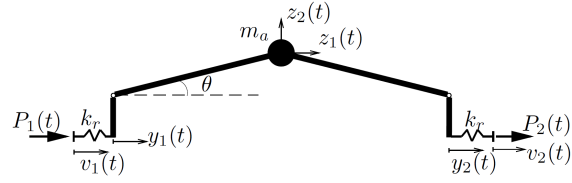}
\caption{LR mechanism with constraint forces.\label{fig:localmechanism}}
\end{figure}

Considering the inertial amplification mechanism in Fig. \ref{fig:mechanismIA}, with the applied constraint forces $P_1$ and $P_2$, the motions $y_1$ and $y_2$ correspond to longitudinal motion of the rod at the points $x_1$ and $x_2$. 
Using Lagrange's equations, the governing equations for the mechanism are found, 
\begin{subequations}
\begin{align}
m_1\ddot{y}_1  - m_2\ddot{y}_2 = P_1 \\
m_1\ddot{y}_2 - m_2\ddot{y}_1  = P_2
\end{align}
\label{eq:constraints}
\end{subequations}

\noindent where  $m_1=\frac{m_a}{4}\left(\cot^2\theta +1 \right)$ and $m_2=\frac{m_a}{4}\left(\cot^2\theta -1 \right)$.
Assuming harmonic motion, $y_j = Y_je^{i\omega t}$, we obtain

\begin{subequations}
\begin{align}
k_1(\omega)Y_1 - k_2(\omega)Y_2 = \widehat{P}_1(\omega) \\
k_1(\omega)Y_2 - k_2(\omega)Y_1 = \widehat{P}_2(\omega) 
\end{align}
\label{eq:constforceia}
\end{subequations}

\noindent were $\widehat{P}_j$ are the frequency domain representations of $P_j$ and the dynamic stiffness parameters $k_j(\omega)$ are defined as:
\begin{equation}
 k_j(\omega) =  -\omega^2m_j \quad , \quad j = 1,2.
\end{equation}
Next, the constraint forces for the local-resonator-type system in Fig. \ref{fig:localsystem} are determined. Considering the isolated mechanism in Fig. \ref{fig:localmechanism}, the constraint forces are applied at the constrained coordinates $v_1$ and $v_2$, while the coordinates $y_1$ and $y_2$ are free.

The governing equations are found by Lagrange's equations,
\begin{subequations}
\begin{align}
&k_r(v_1 - y_1) = P_1 \\
&m_1\ddot{y}_1 +  k_r y_1 - m_2\ddot{y}_2  - k_rv_1 = 0 \\
&m_1\ddot{y}_2 +  k_ry_2 - m_2\ddot{y}_1  - k_rv_2 = 0\\
&k_r(v_2 - y_2) = P_2.
\end{align}
\end{subequations}
Assuming harmonic motion provides the frequency domain constraint forces, which in terms of the constraint coordinates $V_1$ and $V_2$ are

\begin{subequations}
\begin{align}
 k_1^{lr}(\omega)V_1 - k_2^{lr}(\omega) V_2 = \widetilde{P}_1(\omega) \\
k_1^{lr}(\omega)V_2 - k_2^{lr}(\omega)V_1 = \widetilde{P}_2(\omega) 
\end{align}
\end{subequations} 
with the dynamic stiffness parameters defined as
\begin{subequations}
\begin{align}
k_1^{lr}(\omega) &= \frac{\omega^2\omega_{lr,1}^2\left(\omega^2(m_1+m_2)-\omega_{lr,2}^2m_1\right)}{(\omega_{lr,1}^2 - \omega^2)(\omega_{lr,2}^2 - \omega^2)} \\
k_2^{lr}(\omega) &= \frac{-\omega^2\omega_{lr,1}^2\omega_{lr,2}^2m_2}{(\omega_{lr,1}^2 - \omega^2)(\omega_{lr,2}^2 - \omega^2)}\label{eq:lrdynstiffk2}
\end{align}
\label{eq:dynstifflr}
\end{subequations}

\noindent where $\omega_{1,lr}$ and $\omega_{2,lr}$ are the local resonance frequencies defined by,
\begin{subequations}
\begin{align}
\omega_{lr,1} &= \sqrt{\frac{k_r }{m_1+m_2}} = \sqrt{\frac{2k_r}{m_a\cot^2\theta}} \label{eq:lrfreq1} \\
\omega_{lr,2} &= \sqrt{\frac{k_r}{m_1-m_2}} = \sqrt{\frac{2k_r}{m_a}}. \label{eq:lrfreq2}
\end{align}
\label{eq:lrfreq}
\end{subequations}

\noindent Eq. \eqref{eq:lrfreq1} corresponds to the out-of-phase mode while Eq. \eqref{eq:lrfreq2} corresponds to the in-phase mode of the mechanism. In the out-of-phase mode, the mass oscillates in a direction orthogonal to that of the displacement at frequency $\omega_{lr,1}$, thus also benefiting from the inertial amplification effect. The in-phase mode on the other hand corresponds to that of a standard local resonator, since there is no relative acceleration between the attachment points $x_1$ and $x_2$.
It is noted that the dynamic stiffness coefficients for the inertial amplification system are recovered from Eqs. \eqref{eq:dynstifflr} when $k_r \rightarrow \infty$.

\section{Analysis}
In order to characterize the effects of the inertial amplification mechanism, the band structure of an infinite array of hybrid rod-mechanism systems is determined using the transfer matrix method. The method has its origins within electrodynamics and optics, \cite{Mochan1987} but has been widely used within elastic wave propagation.\cite{Esquivel-Sirvent1994,Hussein2006}  The method is briefly described in Section \ref{ssec:transfermatrix}, with a focus on the specific extension required for the particular unit cell considered here. 
Before describing the general transfer-matrix methodology, we consider a simplified unit cell in Section \ref{ssec:receptance} in order to shed light on the band-opening mechanism. 
We do this by a \emph{receptance approach}, \cite{Bishop1979} where we determine the displacement at one end of a single unit cell when applying harmonic forcing at the other end. The anti-resonance frequencies can then be determined as those frequencies with zero receptance for any forcing magnitude (zeros). These anti-resonance frequencies are shown to be the points of maximum attenuation in the infinite system, and are thus relevant quantities for maximum attenuation design.

\subsection{Receptance approach\label{ssec:receptance}}
Consider the simplified rod-mechanism system in Fig. \ref{fig:hybrid} with free boundary conditions, $l_1=l_3 = 0$, and harmonic forcing, $F = F_0e^{i\omega t}$, at $x=0$. 
Both the applied and constraint forces can be included via the boundary conditions to the rod differential equation, Eq. \eqref{eq:bardl}. With harmonic forcing, the linear response will be harmonic, $u(x,t) = \bar{u}(x)e^{i\omega t}$, whereby
\begin{equation}
\bar{u}''(x) + \kappa_b^2\bar{u}(x) = 0 \, , \quad \kappa_b= \frac{\omega}{c_0}
\label{eq:spatialsimple}
\end{equation}

\noindent with wavenumber $\kappa_b$ and wave-speed $c_0 = \sqrt{E/\rho}$ in the homogeneous rod.
The solution to \eqref{eq:spatialsimple} is
\begin{equation}
\bar{u}(x) = \alpha\sin\kappa_b x + \beta \cos \kappa_b x
\label{eq:usol}
\end{equation}

\noindent where the constants $\alpha$ and $\beta$ are determined by the boundary conditions, given by force equilibria at both ends. Utilizing the constitutive relation  $\bar{\sigma}(x) = E\bar{u}'(x)$, the force equilibria yield
\begin{subequations}
\begin{align}
x = 0:& \quad EA\bar{u}'(0) = F_0 - \widehat{P}_1(\omega) \\
x = l:& \quad EA\bar{u}'(l) = -\widehat{P}_2(\omega)
\end{align}
\end{subequations}
\noindent which, when inserting the constraint forces from Eqs. \eqref{eq:constforceia} and the solution $\bar{u}(x)$ from Eq. \eqref{eq:usol}, are expressed in matrix form, 
\begin{equation}
\left[\begin{array}{cc} 1-\hat{k}_2\sin\kappa_b l & \hat{k}_1 - \hat{k}_2\cos\kappa_b l \\ \cos\kappa_b l + \hat{k}_1\sin\kappa_b l & \hat{k}_1\cos\kappa_b - \sin\kappa_b l - \hat{k}_2 \end{array}\right]\textbf{x} = \textbf{f}
\end{equation}
with $\hat{k}_i = k_i(\omega)/(EA\kappa_b)$, $\textbf{x} = [\alpha \; \beta]^T$ and $\textbf{f}~=~[F_0/(EA\kappa_b) \; 0]^T$.
Solving for $\alpha$ and $\beta$ provides the solution for $\bar{u}(x)$,
\begin{widetext}
\begin{equation}
\bar{u}(x) = \frac{\left((\sin\kappa_b l - \hat{k}_1\cos\kappa_b l + \hat{k}_2)\sin\kappa_b x  + \left(\cos\kappa_b l + \hat{k}_1\sin\kappa_b l\right)\cos\kappa_b x\right)}{\sin\kappa_bl\left(1+ \hat{k}_1^2 -\hat{k}_2^2\right)EA\kappa_b}F_0 = H_{0x}(\omega)F_0
\end{equation}
\end{widetext}
where $H_{0x}(\omega)$ is the receptance function.
The displacement at $x=l$ is given by

\begin{equation}
\bar{u}(l) = \frac{\left(1 + \hat{k}_2\sin\kappa_b l\right)}{\sin\kappa_bl\left(1+ \hat{k}_1^2 -\hat{k}_2^2\right)EA\kappa_b}F_0 = H_{0l}(\omega) F_0
\end{equation}

\noindent whereby the anti-resonance frequencies can be determined as the frequencies satisfying $H_{0l}(\omega) = 0$, i.e.,
\begin{equation}
k_2(\omega)\sin\kappa_b l + EA\kappa_b = 0.
\label{eq:ar}
\end{equation}
This transcendental equation is solved  numerically for any desired number of anti-resonance frequencies. The approximation for the first anti-resonance frequency is found in the long-wavelength limit where $\sin\kappa_b l \approx \kappa_b l$, i.e. in the sub-Bragg regime. The approximation $\tilde{\omega}_{a,1}$ is

\begin{equation}
\begin{split}
-& \omega^2m_2\kappa_b l + EA\kappa_b = 0 \Rightarrow  \\
 &\tilde{\omega}_{a,1} = \sqrt{\frac{EA/l}{m_2}} = \sqrt{\frac{EA/l}{m_a(\cot^2\theta - 1)/4}}
\label{eq:approxar}
\end{split}
\end{equation}

\noindent which is essentially the same as the anti-resonance frequency of a discrete system as presented in Ref. \onlinecite{Yilmaz2007} with effective spring stiffness $k = EA/l$. Hence discretizing the rod as a spring-mass system would provide the semi-infinite gap presented in the mentioned reference. 
The added complexity from the rod is illustrated by the higher roots of Eq. \eqref{eq:ar} and will be apparent from the band structures calculated in Section \ref{sec:results}.
\subsection{Transfer matrix method\label{ssec:transfermatrix}}
The transfer matrix method is based on relating the state variables of a system across distances and interfaces, successively creating a matrix product from all the ``sub'' transfer matrices, forming the \emph{cumulative transfer matrix}. 

Consider the hybrid continuous-discrete, rod-mechanism system illustrated in Fig. \ref{fig:hybrid}, where the rod is modelled as a continuum and the mechanism is modelled by discrete elements. 
The transfer matrix for the unit cell is based on the host medium, i.e., the rod, representing the effects of the mechanisms by point force matrices at $x_1$ and $x_2$. 
The state variables for the rod are the longitudinal displacement $u(x,t)$ and the normal stress $\sigma(x,t)$. 
Dividing the system into three layers separated at $x_1$ and $x_2$, the solution for the longitudinal displacement in layer $j$ can be written as a sum of forward and backward travelling waves,

\begin{equation}
u_j(x,t) = \left(B_j^{(+)}e^{i\kappa_bx} + B_j^{(-)}e^{-i\kappa_bx}\right)e^{i\omega t}
\end{equation} 
where $B_j^{(+)}$ and $B_j^{(-)}$ are the amplitudes of the forward and backward travelling waves. Using the linear elastic constitutive relation for the stress, the state variables are expressed as
\begin{equation}
\begin{split}
\textbf{z}_j(x) =& \left[\begin{array}{c}u_j(x) \\ \sigma_j(x)\end{array}\right] = \left[\begin{array}{cc} 1 & 1 \\Z & -Z\end{array}\right]\left[\begin{array}{l}B_j^{(+)}e^{i\kappa_b x} \\[0.1cm] B_j^{(-)}e^{-i\kappa_bx}\end{array}\right] \\
=& \textbf{H}\left[\begin{array}{l}B_j^{(+)}e^{i\kappa_b x} \\[0.1cm] B_j^{(-)}e^{-i\kappa_bx}\end{array}\right]
\label{eq:transferbegin}
\end{split}
\end{equation}
thus defining the \textbf{H}-matrix, where $Z = iE\kappa_b$.
Relating the state variables at either end of a homogeneous layer separated by the distance $l_j$ yields
\begin{equation}
\begin{split}
\textbf{z}_j^R =& \textbf{H}\left[\begin{array}{cc} e^{i\kappa_b l_j} & 0 \\ 0 & e^{-i\kappa_b l_j}\end{array}\right]\left[\begin{array}{l}B_j^{(+)}e^{i\kappa_b x^{j,L}} \\[0.1cm] B_j^{(-)}e^{-i\kappa_bx^{j,L}}\end{array}\right] \\
=& \textbf{H}\textbf{D}_j\left[\begin{array}{l}B_j^{(+)}e^{i\kappa_b x^{j,L}} \\[0.1cm] B_j^{(-)}e^{-i\kappa_bx^{j,L}}\end{array}\right]
\label{eq:phasematrix}
\end{split}
\end{equation}
defining the ``phase-matrix'' $\textbf{D}_j$. The coordinate at the left end of layer $j$ is denoted $x^{j,L}$.
Solving Eq. \eqref{eq:transferbegin} with $\textbf{z}_j = \textbf{z}_j^L$ for the vector of amplitudes and inserting into Eq. \eqref{eq:phasematrix} defines the transfer matrix for layer $j$,
\begin{equation}
\begin{split}
&\textbf{z}_j^R = \textbf{HD}_j\textbf{H}^{-1}\textbf{z}_j^L = \textbf{T}_j\textbf{z}_j^L \\
& \textbf{T}_j = \left[\begin{array}{cc} \cos\kappa_bl_j & \frac{1}{E\kappa_b}\sin\kappa_bl_j \\ -E\kappa_b\sin\kappa_bl_j & \cos\kappa_bl_j\end{array}\right].
\end{split}
\end{equation}

\noindent 
Having defined the transfer matrices $\textbf{T}_j$, $j = 1,2,3$, we turn to the constraint forces at the attachment points of the mechanism. 

We base our derivation of the point force matrices on a frequency domain force equilibrium at the attachment points, considering $x_1$ first
\begin{equation}
A\sigma_2^L = A\sigma_1^R + \widehat{P}_1 = A\sigma_1^R  + k_1(\omega)u_1^R - k_2(\omega)u_2^R.
\end{equation}
It is noted that the force balance at point $x_1$ depends on the displacement at $x_2$. Using the transfer matrix for layer 2, $u_2^R$ is expressed  as
\begin{equation}
u_2^R = u_2^L \cos\kappa_bl_2 +\frac{\sigma_2^L}{E\kappa_b}\sin\kappa_bl_2
\end{equation}
which, along with the continuity requirement $u_1^R = u_2^L$, yields the force equilibrium

\begin{equation}
\begin{split}
\left(A+\frac{k_2(\omega)}{E\kappa_b}\sin\kappa_bl_2\right)&\sigma_2^L = A\sigma_1^R  \\
+ \big(k_1(\omega)&
- k_2(\omega)\cos\kappa_bl_2\big)u_1^R
\end{split}
\end{equation}
whereby the point force matrix, relating the state vector $\textbf{z}_2^L$ to $\textbf{z}_1^R$ can be identified,

\begin{equation}
\begin{split}
\textbf{z}_2^L =& \left[\begin{array}{cc} 1 & 0 \\ \frac{E\kappa_b\left(k_1(\omega)-k_2(\omega)\cos\kappa_bl_2\right)}{AE\kappa_b + k_2(\omega)\sin\kappa_bl_2} & \frac{AE\kappa_b}{AE\kappa_b + k_2(\omega)\sin\kappa_bl_2} \end{array}\right]\textbf{z}_1^R\\ =& \widehat{\textbf{P}}_1\textbf{z}_1^R.
\label{eq:pointforce1}
\end{split}
\end{equation}
Using a similar approach at point $x_2$ provides the point force matrix $\widehat{\textbf{P}}_2$ as

\begin{equation}
\widehat{\textbf{P}}_2 = \left[\begin{array}{cc} 1 & 0 \\ \frac{k_1(\omega)-k_2(\omega)\cos\kappa_bl_2}{A} & 1+ \frac{k_2(\omega)}{AE\kappa_b}\sin\kappa_b l_2\end{array}\right]
\label{eq:pointforce2}
\end{equation}
which allows for relating the state vector at the right end of the unit cell to the state vector at the left end through the cumulative transfer matrix $\textbf{T}$,

\begin{equation}
\textbf{z}_3^R = \textbf{T}_3\widehat{\textbf{P}}_2\textbf{T}_2\widehat{\textbf{P}}_1\textbf{T}_1\textbf{z}_1^L = \textbf{T}\textbf{z}_1^L.
\label{eq:transfermatrix}
\end{equation}
The present framework is fully compatible with the local-resonator-type system described in Section \ref{sec:model} by changing the dynamic stiffness parameters in the point-force matrices to those of Eqs. \eqref{eq:dynstifflr}, rather than those defined by Eqs. \eqref{eq:constforceia}.
Finally, it is noted that when the internal distance $l_2$ approaches zero, the point force matrices for the inertial amplification system approach that of an attached point mass, while the local-resonance point force matrices approach that of an attached local resonator with resonance frequency $\omega_{lr,2} = \sqrt{2k_r/m_a}$, recovering the expected limits.

With the cumulative transfer matrix for a unit cell determined, the Floquet-Bloch theorem for periodic structures, \cite{Bloch1929} is used to relate the state vector at either end through a phase multiplier
\begin{equation}
\textbf{z}_3^R = \textbf{z}_1^Le^{i\kappa l}
\label{eq:floquet}
\end{equation}
where $\kappa = \kappa(\omega)$ is the wavenumber for the periodic material and $l$ is unit-cell length. Combining Eqs. \eqref{eq:transfermatrix} and \eqref{eq:floquet}  yields a frequency-dependent eigenvalue problem in $e^{i\kappa l}$
\begin{equation}
\left(\textbf{T} - e^{i\kappa l}\textbf{I}\right)\textbf{z}_1^L = \textbf{0}
\end{equation}
whereby the band structure of our periodic material system is determined within a $\kappa(\omega)$-formulation.

\section{Band structure\label{sec:results}}
In this section, the band structure is calculated for the systems in Figs. \ref{fig:hybrid} and \ref{fig:localsystem} as well as a standard local resonator configuration, with primary focus on the inertial amplification system. The mechanism is attached to an aluminum rod with Young's modulus $E$, density $\rho$, width $b$, height $h$ and length $l$. These parameters are given in Table \ref{tab:mainbar} along with the equivalent mass and stiffness parameters $m_b$ and $k_b$ and the first natural frequency $\omega_b$. 

\begin{table}[h!t]
\centering
\caption{Parameters of the host rod}
\label{tab:mainbar}
\begin{tabular}{cccc|ccc}
\toprule
$E$ & $\rho$ & $b\times h$ & $l$ & $\omega_b$ & $m_b$ & $k_b$ \\
$[\text{GPa}]$ & $\left[\text{kg/m}^3\right]$ & $\left[\text{m}\right]\times\left[\text{m}\right]$ & [m] & [rad/s] & [kg] & [N/m]  \\[0.2cm]
\hline 
69.8 & 2700 & $0.025 \times 0.025$ & 0.55 & $\frac{\pi}{l}\sqrt{\frac{E}{\rho}}$ & $\rho A l$ & $\frac{EA}{l}$  \\
\toprule
\end{tabular}
\end{table}

\noindent The effect of the primary parameters of the system is investigated, with special attention devoted to gap width and depth. 
We begin by considering the case where the internal length of the mechanism is equal to the unit-cell length, i.e., $l_1 = l_3 = 0$, illustrating the effect of the added mass $m_a$, after which we consider the effect of reducing the mechanism size compared to the unit-cell length. Next we consider the local-resonator-type system and the transition from local resonance to inertial amplification, after which we compare the performance of the IA system to that of a standard local resonator system where the resonator is attached to the rod at a point within the unit cell. 

\subsection{Band structure and anti-resonance frequencies\label{ssec:numreference}}
We begin by considering a reference case, with the rod parameters in Table \ref{tab:mainbar} and the mechanism parameters given in Table \ref{tab:mechanism}.

\begin{table}[h!t]
\centering
\caption{Reference mechanism parameters}
\label{tab:mechanism}
\begin{tabular}{ccc}
\toprule
$ m_a$ & $\theta$ & $l_2$\\
\hline
$m_b/10$ & $\pi/18$ & $l$ \\
\toprule
\end{tabular}
\end{table}

\begin{figure} 
	\centering
		\includegraphics[width=0.43\textwidth]{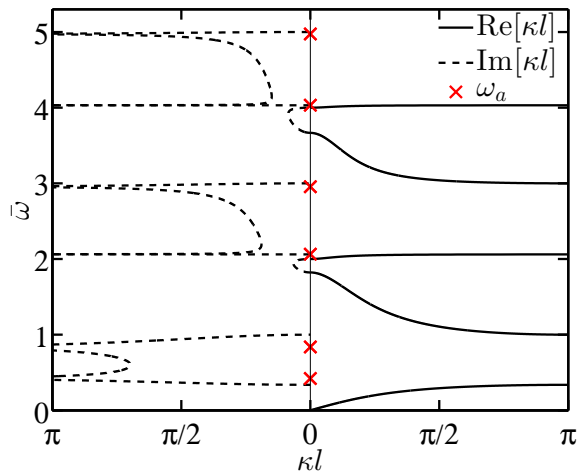}
	\caption{Band-structure for reference case.\label{fig:reference}}
\end{figure}

\noindent For the reference case, the anti-resonance frequencies are predicted by numerically solving Eq. \eqref{eq:ar}, while the band structure is calculated using the transfer-matrix method described in Sec. \ref{ssec:transfermatrix}.
Figure \ref{fig:reference} shows the band structure with the anti-resonance frequencies indicated by red crosses, in the normalized frequency range $\bar{\omega} = \omega/\omega_b$.

It is clear that the solution of Eq. \eqref{eq:ar} predicts the peak values of the imaginary part of the wave number. 
A feature unique to inertial amplification seen in the band structure in Fig. \ref{fig:reference} is the appearance of two peaks within the odd-numbered gaps, that are noted to have merged with the odd-numbered Bragg gaps with upper limits at $\bar{\omega} = 1,3,5,\cdots$. 
This double-peak behaviour is behaviour is previously observed in Ref. \onlinecite{Yilmaz2006,Acar2013} however in the frequency response functions of finite structures rather than the band structure of the system. 
The reason that the inertial amplification gaps and Bragg gaps merge is the fact that the mechanism is attached to the ends of the unit cell, hence the wavelengths most significant for activation of the inertial amplification mechanism are the same as those relevant for Bragg scattering. 
The number of attenuation peaks within a band gap is determined by the number of anti-resonance frequencies separating the two wave modes on either side of the gap. Hence when two resonance frequencies (poles) are separated by more than one anti-resonance frequency (zero), multiple attenuation peaks will occur. 
The low-pass filters shown in Ref. \onlinecite{Yilmaz2006} are designed such that all the anti-resonance frequencies are larger than the largest resonance frequency, whereby the multiple peaks are all found above the filter frequency.
It is further noted that the even gaps are of Bragg-type, however they exhibit an asymmetry, being distorted towards the double-peak gaps which is also common when Bragg gaps are close to the peak of local-resonance gaps. \cite{Xiao2011,Liu2012} 

When considering the band structure for increasing frequency, it appears that for higher band gaps, the anti-resonance frequencies get further separated, increasing the gap width at the cost of decreasing the gap depth. Furthermore, as the gap widens, the anti-resonance frequencies approaches the gap limits. This indicates that the anti-resonance frequencies can be used as a design-parameter for gap width. This is investigated further in the Sec. \ref{ssec:num_attachment}, where it is shown that this is only strictly true for ``full-length'' mechanisms, i.e., when $l_1 = l_3 = 0$.

\subsection{Mass variation} 
\begin{figure*} 
	\centering
	\includegraphics[width=0.92\textwidth]{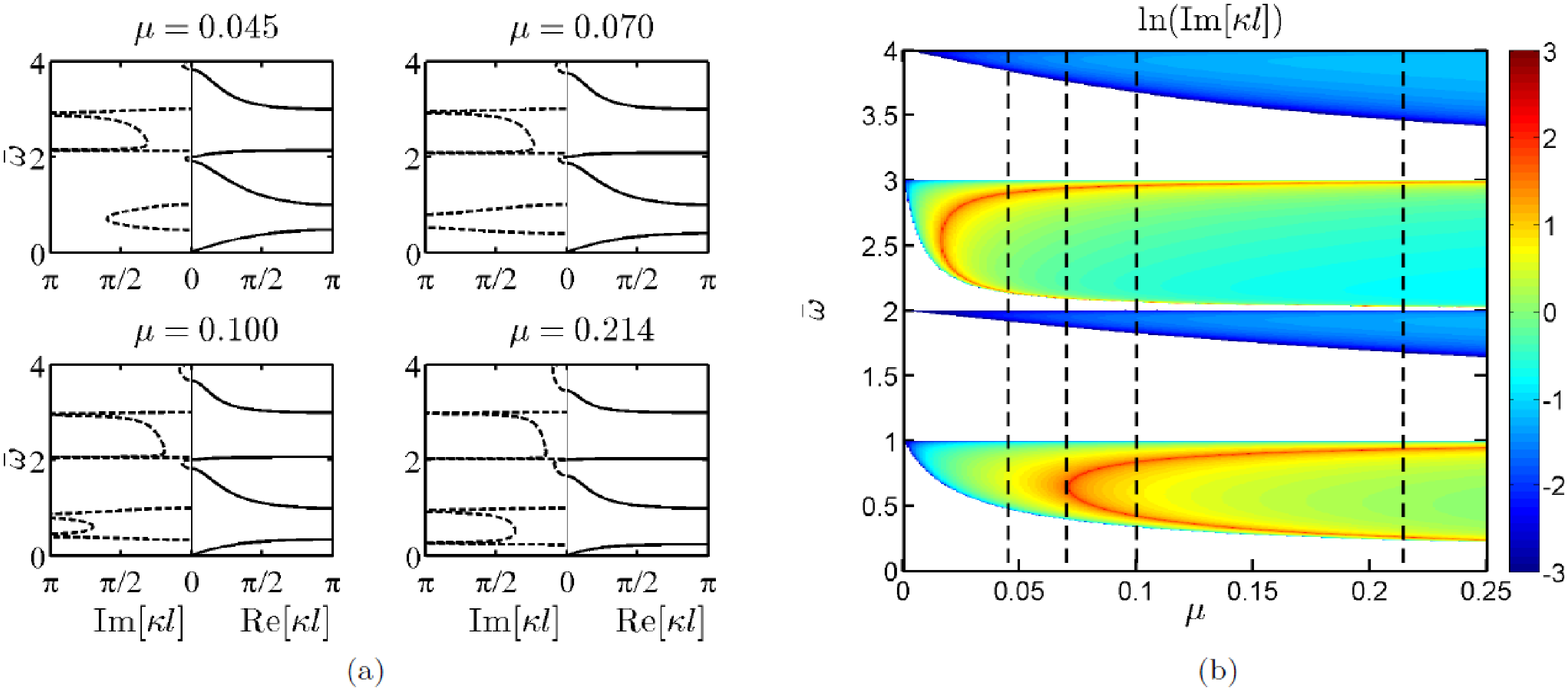}		
	\caption{ Effect of amplification mass for full length mechanism.(a) Band structures. (b) Gap variation.\label{fig:massvariation_FullLength}}
\end{figure*}

Consider the model shown in Fig. \ref{fig:hybrid} with $l_1 = l_3 = 0$. We investigate the effect of the attached mass $m_a$ for the general rod parameters in Table \ref{tab:mainbar} and the mechanism parameters in Table \ref{tab:mechanism}.
It is expected that increasing the attached mass will increase the gap width as well as decrease the frequency range. 
These effects are illustrated for distinct values of the mass ratio $\mu = m_a/m_b$ in Fig. \ref{fig:massvariation_FullLength}(a) as well as by the \emph{attenuation profile} in Fig. \ref{fig:massvariation_FullLength}(b) which in this paper refers to a contour plot of the natural logarithm of the imaginary part of the wave number. 
The four band structures in Fig. \ref{fig:massvariation_FullLength}(a) are for the $\mu$-values indicated by the four vertical dashed lines in the attenuation profile in Fig. \ref{fig:massvariation_FullLength}(b). 
It is noted that for the mass ratio $\mu = 4.5\%$, the first gap is purely of Bragg-type since the inertial forces generated by the mechanism are not sufficient to generate anti-resonance frequencies, i.e., to fulfil Eq. \eqref{eq:ar}.
As $\mu$ is increased, the inertial forces are increased (by a factor of $m_a$), and $\mu = 7\%$ is seen to be the limiting value for generating true anti-resonance frequencies. At this value, we see a single anti-resonance frequency which branches into two distinct anti-resonance frequencies as $\mu$ is further increased. 

Aside from this double-peak behaviour, the width of the first gap is rather wide for relatively low mass-ratios, e.g., for a mass ratio of 10\% we have a normalized first gap width of
\begin{equation}
\Delta \bar{\omega}_1 = \frac{\bar{\omega}_1^u - \bar{\omega}_1^l}{\bar{\omega}_1^c} = 98.8 \%
\end{equation} 
which, compared to gaps obtained by standard LR effects, is rather wide, as will be shown in Sec. \ref{ssec:standardlr}.
The logarithmic scale in Fig. \ref{fig:massvariation_FullLength}(b) illustrates the order of magnitude of the gap depth. Between the peak values at the anti-resonances, the attenuation is seen to stay at a relatively high level, a level that decreases with gap-width as observed for increasing gap number in Section \ref{ssec:numreference}.

\subsection{Effect of attachment points\label{ssec:num_attachment}}
\begin{figure*} 
	\centering
\includegraphics[width=0.92\textwidth]{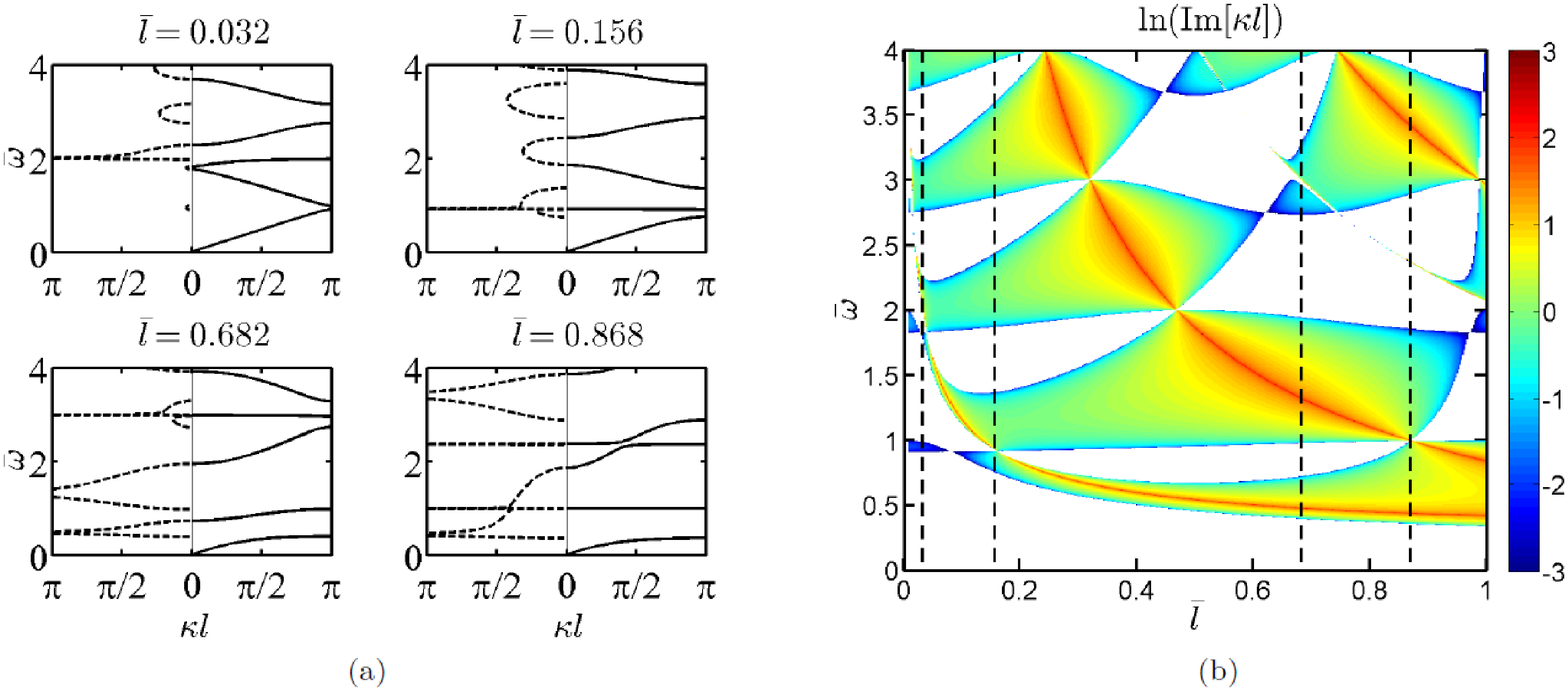}		
		\caption{ Effect of decreasing $\bar{l} = l_2/l$. $\theta =\pi/18$, $\mu = 0.1$. (a) Band structures. (b) Gap variation.\label{fig:attachmentvariation}}
\end{figure*}

Investigating the effect of the relative mechanism length $\bar{l}=l_2/l$, it is expected that the performance metrics, i.e., the gap width and  depth, of the mechanism will decrease with a smaller internal length due to the decreased lever-arm for the mass. It is not a trivial investigation however, since decreasing the mechanism size can provide new ways for gaps to hybridize, e.g. the ``even''-numbered wave-modes will be affected by the mechanism. 
Figure \ref{fig:attachmentvariation} illustrates the effect of varying the relative attachment length, $\bar{l}=l_2/l$, with four band diagrams in Fig. \ref{fig:attachmentvariation}(a) and the attenuation profile in Fig. \ref{fig:attachmentvariation}(b), for the general rod parameters in Table \ref{tab:mainbar} and the mechanism parameters from Table \ref{tab:mechanism}.
The four band structures in Fig. \ref{fig:attachmentvariation}(a) correspond to the vertical dashed lines in the attenuation profile in Fig.~\ref{fig:attachmentvariation}(b). 
Considering the attenuation profile first, the first striking feature is the multitude of coalescence points between the gaps, where the anti-resonance frequencies jump from one gap to another. 
At these coalescence points, the gap width is effectively increased to the sum of the two coalescent gaps, e.g. at $\bar{l} = 86.8\%$, we get an effective, normalized gap width of $\Delta\bar{\omega} = 135.5 \%$.
The corresponding band-structure illustrates that the two gaps do not coalesce completely, rather there is a standing-wave mode separating the two gaps. Hence the second anti-resonance frequency is in fact larger than the second resonance frequency, whereby the first gap only contains one anti-resonance frequency.
This indicates the presence of a \emph{critical region} between the gap limits, especially for practical designs where design inaccuracies will make an exact realization of $\bar{l} = 86.8\%$ difficult. 
The three remaining band structures, at $\bar{l} = 3.2 \%$, $15.6 \%$ and 68.2 \% respectively, illustrate various effects. 
The first band structure illustrates the limiting behaviour towards the pure Bragg gaps from periodically attached masses when $\bar{l} \rightarrow 0$. It is noted that this transition occurs earlier for lower gaps, due to the shorter wavelengths of the higher gaps. 
The second band structure illustrates the first coalescence point between the first and second gap, the effect of which is not quite as powerful as for the $\bar{l}=86.8 \%$ case, however this can be explained be the smaller lever ratio, thus providing smaller inertial forces. The third band-structure for $\bar{l} = 68.2 \%$ illustrate the first two distinct gaps, as well as a Bragg-type gap which is virtually unaffected by the anti-resonance frequency very close to it.
\subsection{Transition from  local resonance\label{ssec:devicelr}}
\begin{figure*} 
\centering
\includegraphics[width=0.92\textwidth]{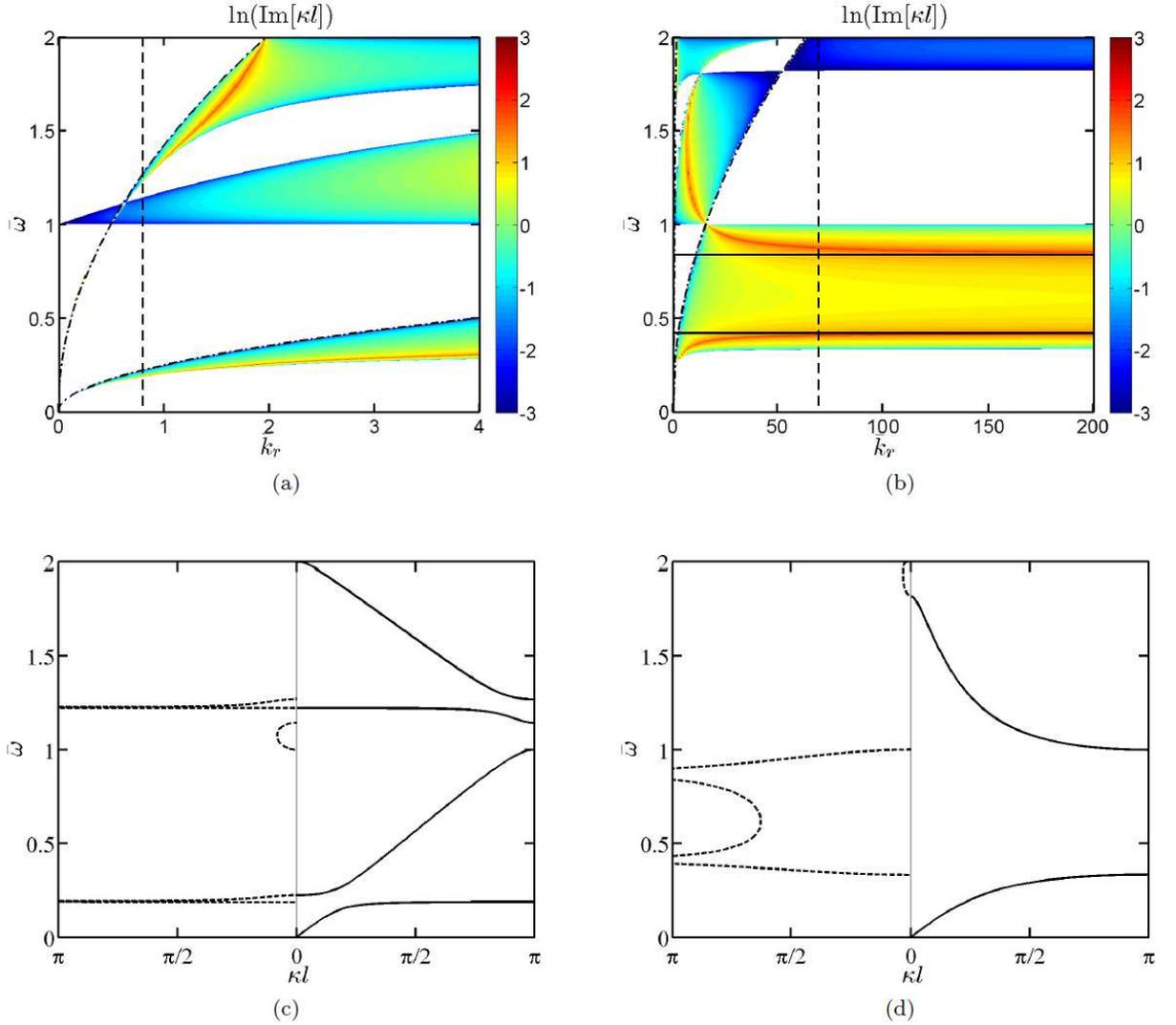}
\caption{ Effect of connection stiffness $\bar{k}_r = k_r/k_b$. $\theta = \pi/18$, $\mu = 0.1$. (a) Attenuation profile in a low range. (b) Attenuation profile in a wide range. (c) Band structure for $\bar{k}_r = 0.8$ (local resonance). (d) Band structure for $\bar{k}_r = 69.3$ (inertial amplification).\label{fig:lrtransition}}
\end{figure*}

In this section, we illustrate how the behaviour of the local-resonator-type system illustrated in Fig. \ref{fig:localsystem} tends toward that of the pure inertial amplification mechanism as the connection stiffness increases, $k_r \rightarrow \infty$. 
The local resonance frequencies seen in Eqs. \eqref{eq:lrfreq} are for the two modes supported by the mechanism, i.e., where the mechanism ends move out-of-phase and in-phase, respectively. 
Figure \ref{fig:lrtransition} shows the attenuation profile for increasing the relative connection stiffness $\bar{k}_r = k_r/k_b$, both in a low range to illustrate the local-resonance effect and a larger range to illustrate the limiting behaviour. 
The dash-dot lines are the resonance frequencies for the local resonance-system predicted by Eqs. \eqref{eq:lrfreq}, noted to follow gap limits, rather than peak attenuation points, as is the case for a standard local resonator attached at a point.
This has a rather simple physical explanation, when considering the working principle of the two types of systems. At their specific resonance frequencies, both resonators are maximally activated, but the fundamental difference lies in the number of attachment points. 
While the standard local resonator will ``take out'' some of the mechanical energy of the travelling wave, the two-terminal system will work as a path of travel for the wave, whereby the mechanical energy is not ``deposited'' as in the standard local resonator case. 
Hence, the peak attenuation frequency for the system including connection flexibility should be predicted from the anti-resonance equation, Eq. \eqref{eq:ar}, but with the dynamic stiffness coefficient given by Eq. \eqref{eq:lrdynstiffk2}, $k_2 = k_2^{lr}$.
The black horizontal lines in Fig. \ref{fig:lrtransition}(b) are the anti-resonance frequencies predicted by Eq. \eqref{eq:ar} using the IA dynamic stiffness coefficient, $k_2 = -\omega^2m_2$, which the peak attenuation lines can be seen to converge towards, starting at the lower frequencies.
Hence, as the connection stiffness increases, the lower gaps become dominated by inertial amplification effects, followed by the higher gaps. 
This is illustrated further by considering the band structures at corresponding to the vertical dash-dot lines in Figs. \ref{fig:lrtransition}(a) and \ref{fig:lrtransition}(b). 
In the low-stiffness limit in Fig. \ref{fig:lrtransition}(c) we see a behaviour that is similar to what is observed for classical local resonance, however we know that the local resonance frequency actually represents the upper limit of the gap. The peak attenuation frequency however is the anti-resonance frequency, where energy \emph{cannot} be transferred through the mechanism. In the high-stiffness limit in Fig. \ref{fig:lrtransition}(d) we observe the ideal inertial amplification behaviour, where we have two anti-resonance frequencies separating the first and second wave-mode, leading to the appearance of the characteristic double peak.

\subsection{Comparison to a standard local resonator\label{ssec:standardlr}}
\begin{figure*} 
	\centering
\includegraphics[width=0.92\textwidth]{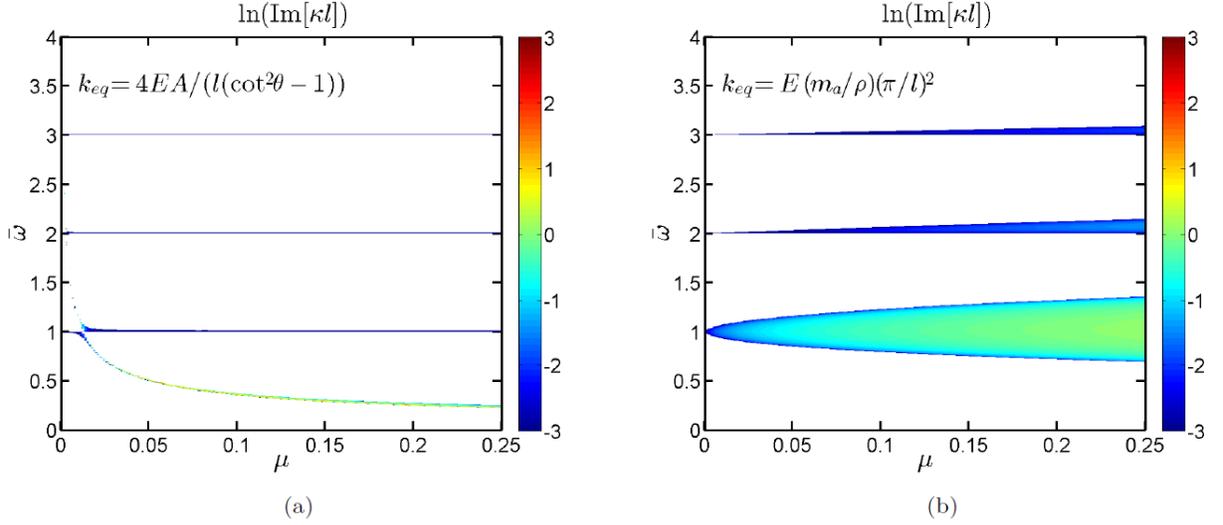}
	\caption{ Effect of mass ratio $\mu$ in a standard local resonance configuration. (a) Stiffness tuning for similar first peak attenuation frequency. (b) Stiffness tuning for local-resonance-Bragg coalescence.\label{fig:mucompare_lr}}
\end{figure*}

The performance metrics are now investigated for a rod with periodic, point-wise attached local resonators, in order to provide a direct performance comparison with the proposed inertial amplification system.
The transfer matrix of such a rod is known and given in, e.g., Ref. \onlinecite{Khajehtourian2014}. One strength of the classical local resonator configuration is the ability to create \emph{any} anti-resonance frequency, given complete freedom in the design variables. 
This is equally possible for the inertial amplification configuration proposed here, which is evident from the anti-resonance equation, Eq. \eqref{eq:ar}. Hence the local resonator offers no design flexibility over the inertial amplification mechanism in this regard. 
A primary parameter for both the inertial-amplification and local-resonator systems is the added mass $m_a$. The effect of the added mass is illustrated for the local-resonator system for two tuning cases of local resonator stiffness, $k_{eq}$. 
One is where the local-resonance frequency is tuned to the approximate first anti-resonance frequency for the inertial amplification system, $\tilde{\omega}_{1,a}$ from Eq. \eqref{eq:approxar}, in order to attain the same peak attenuation frequency for the two systems. 
The second case is where the local resonator system is tuned to allow the local-resonance gap to coalesce with the first Bragg gap-since it is known that this is where local resonance gaps achieve maximal width.\cite{Xiao2011,Xiao2012,Khajehtourian2014} Fig. \ref{fig:mucompare_lr} illustrates the effect of increasing the mass ratio $\mu$ for the two cases. 
The attenuation profile seen in Fig. \ref{fig:mucompare_lr}(a) illustrates how the local-resonance gap decreases in frequency range with increasing mass, while the Bragg gaps stay centred around the same  frequency with mass increase.
Comparing to the inertial amplification system for $l_1 = l_3 = 0$, the inertial amplification system is seen to perform better in terms of both gap width and depth, see Fig. \ref{fig:massvariation_FullLength}.
The attenuation profile for local-resonance-Bragg coalescence in Fig. \ref{fig:mucompare_lr}(b) shows superior gap width than seen for the local-resonance attenuation profile in Fig. \ref{fig:mucompare_lr}(a), however the lowest and widest gap is centred around the first Bragg-gap frequency, thus not providing any low-frequency attenuation. 
Furthermore, the inertial amplification configuration still offers a wider and deeper gap, while being lower in frequency.

\subsection{Effect of unit-cell size}
\begin{figure} 
\centering
\includegraphics[width=0.46\textwidth]{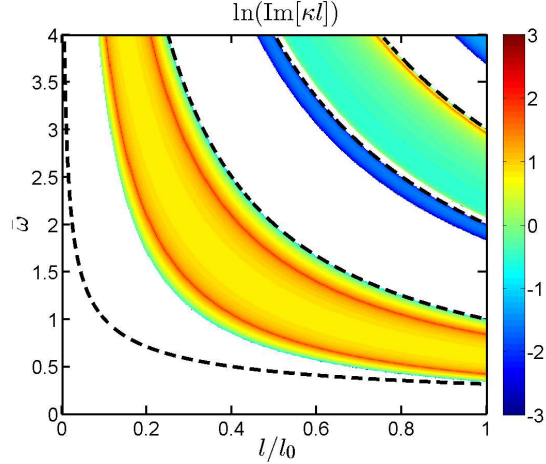}
\caption{ Dependence on unit-cell size. Local resonator: dashed lines, Inertial Amplifier: colored regions.}
\label{fig:unitcellsize}
\end{figure}

Now the attenuation profile for the inertial amplification system is compared to that of a local resonance system when the unit-cell size is decreased.
This is done to examine the practical limitation of the inertial amplification system, given that an internal length is needed to generate the enhanced inertia to cancel the elastic forces in the rod. This limitation is similar to the one investigated in Sec. \ref{ssec:num_attachment}, concerning the internal length between attachment points, $\bar{l}$, however now we look into the unit cell size $l$ rather than relative internal length. 
The standard local resonance system does not need any internal length, and as such, can operate unaltered as the unit-cell size goes towards zero. 
Figure \ref{fig:unitcellsize} compares the attenuation profile for the two systems, when decreasing the unit cell size $l/l_0$ for the parameters seen in Table \ref{tab:sizepar}, where both $\mu$ and $\bar{k}_{eq}$ are based on the original length $l_0$.
\begin{table}[h!t]
\centering
\caption{Parameters for unit-cell size investigation}
\label{tab:sizepar}
\begin{tabular}{ccccc}
\toprule
$l_0$ [m] & $\mu$  & $\theta$ [rad] & $\bar{k}_{eq}$ & $l_1 = l_3$ \\[0.1cm]
0.55 & $0.1$ & $\pi/18$ & $0.1$ & 0 \\
\toprule
\end{tabular}
\end{table}
The dashed lines represent the gap frequencies for the local-resonator system. The first line represents the local resonance frequency $\omega_{eq} = \sqrt{k_{eq}/m_a}$  where $k_{eq}$ is kept constant, while $m_a$ decreases with unit cell size as $\mu$ is kept constant. The subsequent lines represent the Bragg-gap limits $\omega_\text{Bragg} = j\omega_b$, $j = 1,2,3,\cdots$. For this particular mass ratio $\mu = 10\%$, the actual gaps are very narrow. Thus they are covered entirely by the line representation in Fig. \ref{fig:unitcellsize}.

Either way, the conclusion is that for realistic and comparable parameters, the local-resonance system can provide wave attenuation at lower frequencies than the inertial-amplification system, as the unit-cell size decreases. This is a clear advantage of the classical locally resonant systems from the point of view of constraints on unit-cell size.

\section{Finite structures\label{sec:finite}}
In this section the band-structure results from Section \ref{sec:results} are compared to the transmissibility for finite systems comprising a certain number of unit cells, illustrating that the material results are representative in a structural setting as well. 
The transmissibilities are calculated from a finite-element model of the continuous-discrete system in Fig. \ref{fig:hybrid}, using standard linear elements to discretize the continuous rod, with stiffness and mass matrix contributions from the mechanism at appropriate nodes. The simple FE model is compared to an FE model created in the commercial software ABAQUS, using 3D beam elements.

\subsection{Transmissibility gaps}
\begin{figure} 
\centering
\includegraphics[width=0.5\textwidth]{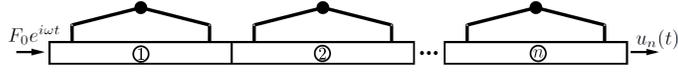}
\caption{Finite array\label{fig:finitearray}}
\end{figure}

Using the FE implementation to model the 1D finite array illustrated in Fig. \ref{fig:finitearray} with the number of unit cells denoted $n$, the transmissibility is expressed as the natural logarithm to the ratio between the output and input displacement divided by the number of unit cells. Hence the transmissibility expresses the wave propagation/decay per unit cell, and is thus comparable to the dispersion curves.

\begin{figure*} 
\centering
\includegraphics[width = 0.85\textwidth]{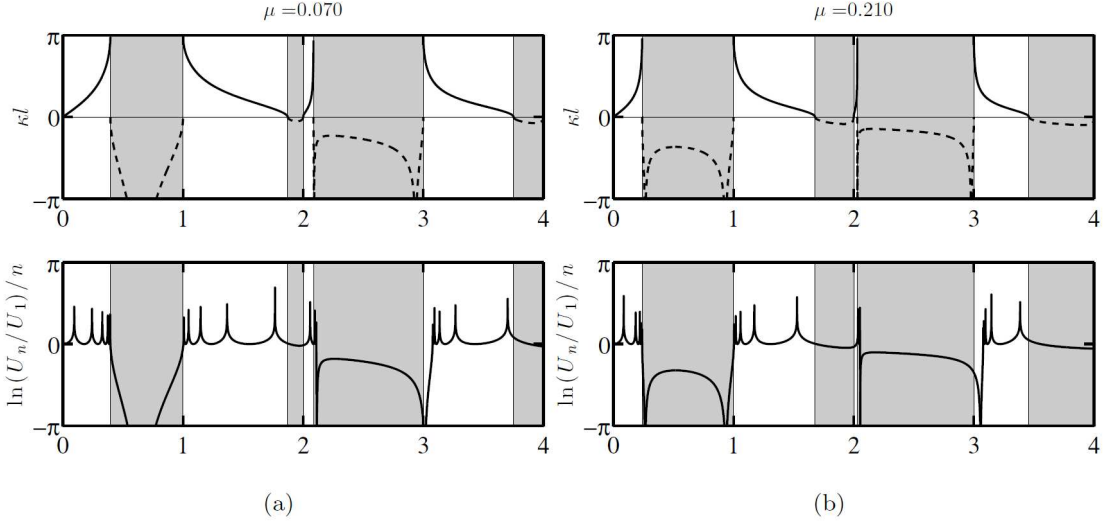}
\caption{Comparison of band structures to transmissibilities for $\bar{l}=1$. (a) Single peak behaviour. (b) Double peak behaviour.\label{fig:femcomparefull}}
\end{figure*}

The transmissibilities are calculated for the same aluminum rod as considered in Section \ref{sec:results}, using the same rod parameters from Table \ref{tab:mainbar}. We consider the case where the relative mechanism length is equal to the full unit cell length, i.e., $\bar{l} = 1$. 
Figure \ref{fig:femcomparefull} compares the band structures from the infinite systems to the transmissibilities calculated for $n=5$ and $F_0 = 10$ N for two values of the mass-ratio $\mu$, illustrating the branching point at $\mu = 7\%$ and the double-dip at $\mu = 21\%$ corresponding to the single- and double-peak in the attenuation profiles, respectively. 
The grey areas correspond to band gaps predicted by the infinite model. It is noted that some boundary effects exist, i.e., the finite systems have resonances within band-gaps, which is due to the symmetry breaking of the system. \cite{Djafari-Rouhani1983,Jensen2003}
In spite of the boundary effects, the comparison shows that the infinite-system properties carry over to the finite case, and perhaps equally important, that within the band gaps, the curves for the imaginary part of the wavenumber and the transmissibility have similar shapes and magnitudes. 
This is expected from the exponential decaying behaviour of waves within the band-gaps, but it does illustrate the design potential for finite structures by just considering the shape of the band-structure within the gaps calculated for infinite structures.

\subsection{ABAQUS verification}
\begin{figure} 
	\centering
	\includegraphics[width=0.5\textwidth]{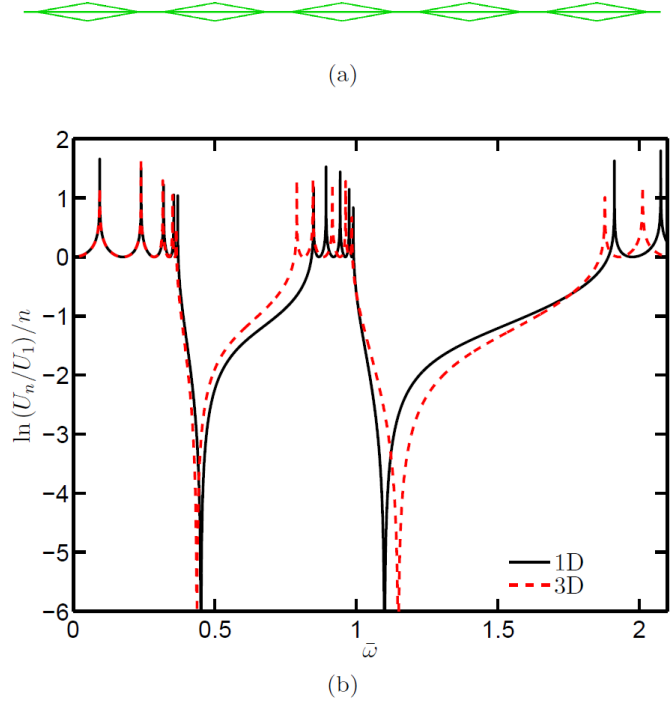} 
	\caption{ (a) ABAQUS Model. (b)Comparison of 3D to 1D model. $\bar{l} = 0.8$, $\mu = 0.1$\label{fig:ABQcomparison_doublerigid_mu_10}}
\end{figure}

The FE implementation of the hybrid rod-mechanism system is tested against an implementation of the finite system in the commercial FE-software ABAQUS. 
The ABAQUS model is created as a 3D deformable 'wire' model, using three-dimensional beam elements for both rod and mechanism. The mechanisms are distributed both above and below the main structure to have equal but opposite transverse force components from the mechanisms. This is necessary to avoid bending phenomena, and may easily be implemented in an experimental setting as well.
The rigid connecting links are modelled by assigning very large Young's modulus and very low density to the elements. The ideal connections are modelled using translatory constraints to connect the rigid connectors to the bar. Hence the ABAQUS model is used to illustrate  the phenomena in a finite setting without obstructing the results with, for the present purpose, unnecessary complexities. Indeed, it is a subject of a future research paper to investigate more realistic models of the physical configuration in Fig. \ref{fig:1D} both numerically and experimentally. 
The ABAQUS model is created with the general rod parameters seen in Table \ref{tab:mainbar} and the general mechanism parameters $\theta = \pi/18$ and $\bar{l} = 0.8$. An illustration of the created ABAQUS model is seen in Fig. \ref{fig:ABQcomparison_doublerigid_mu_10}(a).
		%
Figure \ref{fig:ABQcomparison_doublerigid_mu_10}(b) shows the comparison between the transmissibilities calculated by the 1D FE implementation of the rod-mechanism system and the 3D FE implementation in ABAQUS, respectively, for the mass ratio $\mu = 10\%$. 

\begin{figure*} 
\centering
\includegraphics[width=1\textwidth]{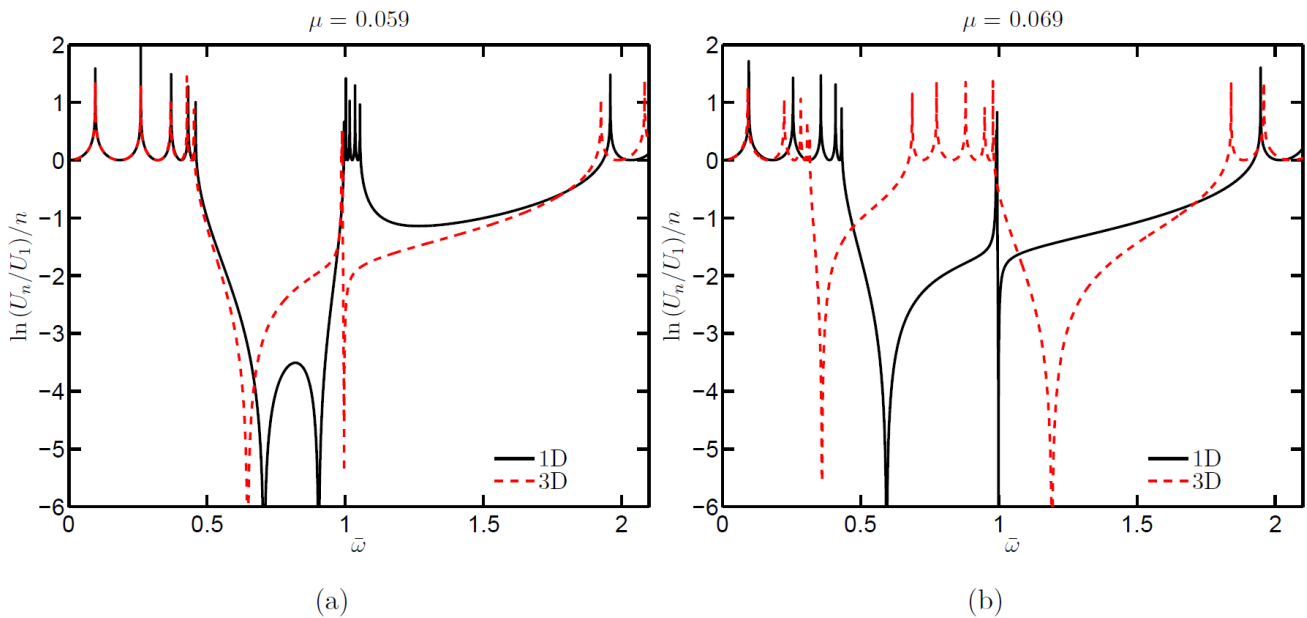}
\caption{ Transmissibility comparison for coalescence in 1D and 3D models for (a)$\mu = 0.059$ and (b) $\mu = 0.069$. \label{fig:ABQcomparison_doublerigid_coalescence}}	
\end{figure*}

Comparing both maximum attenuation frequencies and gap limits, the transmissibility predicted by ABAQUS matches rather well, especially for the first gap. 
Concerning the deviations in the transmissibilities, it is worth considering the case of gap-coalescence which, as seen in Fig. \ref{fig:attachmentvariation} is a rather ``singular'' phenomena. 
As expected, using the ``coalescence-parameters'' predicted by the analytical model does not cause the gaps to coalesce in the 3D model. Figure \ref{fig:ABQcomparison_doublerigid_coalescence} shows the transmissibility comparison for the analytically predicted coalescence-parameters and the ones found by inverse analysis in ABAQUS. 
This pass band could be detrimental for design if not taken into account, since the resonances are so closely spaced. Hence, designing for gap-coalescence should be done with care, as mentioned in Sec. \ref{ssec:num_attachment}.

\section{Conclusions}
We have investigated the wave characteristics of a continuous rod with a periodically attached inertial amplification mechanism. 
The inertial amplification mechanism, which is based on the same physical principles as the classical inerter, creates band gaps within the dispersion curves of the underlying continuous rod.
The gap-opening mechanism is based on an enhanced inertial force generated between two points in the continuum, proportional to the relative acceleration between these two points. 
An inertial amplification mechanism has been used previously as a core building block for the generation of a lattice medium, rather than serve as a light attachment to a continuous structure. 
Several prominent effects are featured in the emerging band structure of the hybrid rod-mechanism configuration.
The anti-resonance frequencies are governed by both mechanism- and rod-parameters, hence rather than a single anti-resonance frequency we see an infinite number, which can be predicted for a simple choice of unit-cell parameters. 
For the same choice, we illustrate the presence of multiple attenuation peaks within the same gap. 
Furthermore, when generalizing the parameters of the unit cell, we observe that the anti-resonance frequencies can jump between gaps whereby double-peak behaviour cannot be guaranteed for all parameters. 
At the specific values of the anti-resonance jump, band-gap coalescence emerges providing a very wide and deep contiguous gap. This gap, however, is rather sensitive to design and modelling inaccuracies.

\begin{figure*} 
\centering
\includegraphics[width=0.9\textwidth]{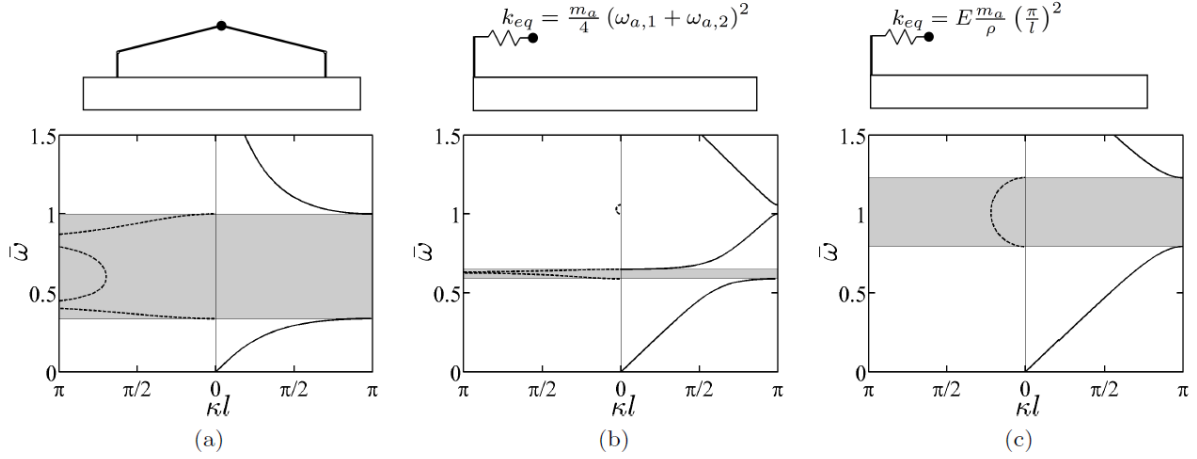}
\caption{Various band-gap opening mechanisms for constant mass ratio $\mu = 10 \%$. (a) Inertial amplification. (b) Local resonance. (c) Bragg scattering.\label{fig:concluding_performance}}
\end{figure*}

\begin{figure*} 
\centering
\includegraphics[width=0.63\textwidth]{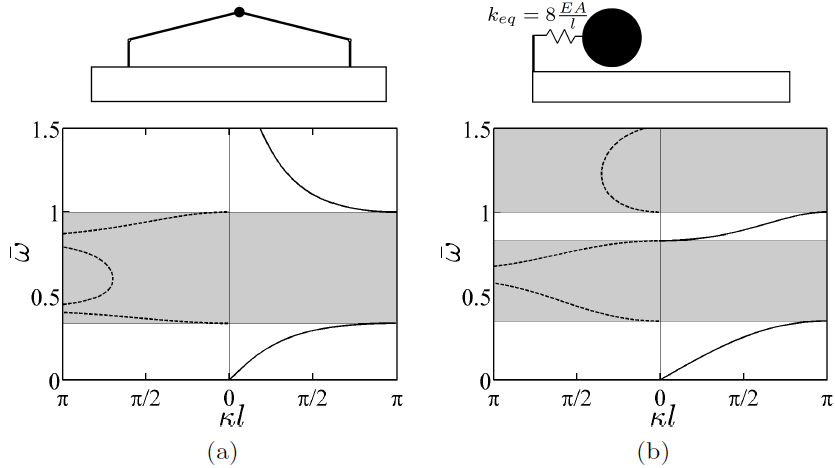}
\caption{Band structure for (a) proposed inertially amplified rod with $\mu = 10\%$ and (b) standard locally resonant rod with $\mu = 204\%$. The systems have almost equal-sized gaps despite a factor of 20 difference in added mass size.\label{fig:concluding_performance2}}
\end{figure*}

In addition to these intriguing effects, we demonstrate how attaching an inertial amplification mechanism to a continuous structure is superior to attaching a classical local resonator in that the former produces much larger gaps for the same amount of added mass. 
Figure \ref{fig:concluding_performance} compares the band structure of the proposed inertial amplification system to those of a classical local resonator configuration for two different tunings of the local resonator stiffness $k_{eq}$. 
Figures \ref{fig:concluding_performance}(b) and \ref{fig:concluding_performance}(c) represent two cases of stiffness tuning that provide a locally resonant band gap (with equal central frequency) and a Bragg coalescence gap, respectively. 
The central frequency is determined by solving Eq. \eqref{eq:ar} numerically for the first two roots, $\omega_{a,1}$ and $\omega_{a,2}$.
The comparison illustrates that when the same mass is used, the proposed concept achieves a first gap that is much wider than what is obtainable by the classical local resonator configuration, irrespective of the stiffness tuning for the local resonator.
In order to obtain comparable performance in terms of band-gap width for the classical local-resonator system, the added mass $m_a$ should be increased significantly. Figure \ref{fig:concluding_performance2} compares similar gap widths for an inertial amplification system and a local resonance system respectively. 
From the figure, we see that the inertial amplification system is superior in terms of the magnitude of added mass, as the local resonance system requires an approximately twenty times heavier mass to obtain a comparable band-gap width (a mass that is more than two times as heavy as the rod it is attached to). 
 The classical local resonator configuration, on the other hand, faces less constraints on unit-cell size as demonstrated in Fig. \ref{fig:unitcellsize}.

The presented concept of an inertially amplified continuous structure opens a new promising avenue of band-gap design.
Potentially it could be extended to surfaces of more complex structures such as plates, shells and membranes leading to a general surface-coating design paradigm for wave attenuation in structures.
Steps toward achieving this goal include a generalization of the formulation to admitting transverse vibrations, incorporation of frictional stiffness and damping in the bearings of the mechanism, and generalization to two dimensions.

\begin{acknowledgments}
Niels M. M. Frandsen and Jakob S. Jensen were supported by ERC starting grant no. 279529 INNODYN. Osama R. Bilal and Mahmoud I. Hussein were supported by the National Science Foundation grant no. 1131802. Further, N.M.M.F would like to extend his thanks to the foundations: COWIfonden, Augustinusfonden, Hede Nielsen Fonden, Ingeni{\o}r Alexandre Haymnan og Hustrus Fond, Oticon Fonden and Otto M{\o}nsted Fonden. 
\end{acknowledgments}

\appendix*

\section{Kinematic derivation\label{app:kinematics}}
\begin{figure*} 
\centering
\includegraphics[width = 0.6\textwidth]{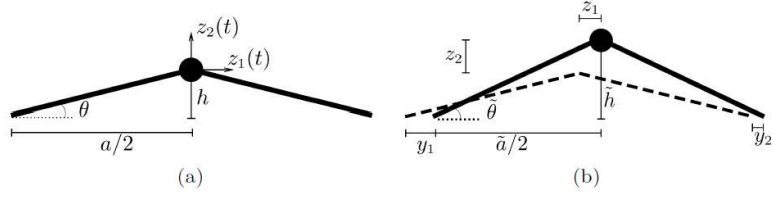}
\caption{Top triangle kinematics. (a) Undeformed triangle. (b) Deformed triangle.\label{fig:topmech_app}}
\end{figure*}

Considering the top part of the deformed mechanism as illustrated in Fig. \ref{fig:topmech_app}, the motions $z_1$ and $z_2$ can be determined in terms of $y_1$, $y_2$ and $\theta$.
The horizontal motion $z_1$ is by geometric consideration determined as $z_1 = (y_1+y_2)/2$. The vertical motion is given by the difference $z_2 = \tilde{h}-h$. It turns out to be convenient to consider the difference of the squared triangle heights:
\begin{widetext}
\begin{equation}
\tilde{h}^2 - h^2 = \left(l^2 - \left(\frac{\tilde{a}}{2}\right)^2\right) - \left(l^2 - \left(\frac{a}{2}\right)^2\right) =  \frac{a^2}{4} -\frac{\left(a+y_2-y_1\right)^2}{4}  = \frac{1}{4}\left(2a(y_2-y_1) - (y_2-y_1)^2\right).
\end{equation}
\end{widetext}
Expressing the difference in squared heights as $\tilde{h}^2-h^2 = (\tilde{h}+h)(\tilde{h}-h) = (2h+z_2)z_2$ and using a geometric relation for $h$ leads to

\begin{equation}
(z_2+a\tan\theta)z_2 = \frac{1}{4}\left(2a(y_2-y_1) - (y_2-y_1)^2\right)
\end{equation}
which provides a quadratic equation in $z_2$. The equation
%
is made explicit by assuming small displacements, such that $(y_2-y_1)^2 << 2a(y_2-y_1)$ and $z_2 << a\tan\theta = h$, whereby the linearized kinematics of the mechanism has been determined as

\begin{subequations}
\begin{align}
z_1 &= \frac{1}{2}(y_2+y_1) \\
z_2 &= \frac{1}{2}\cot\theta(y_2-y_1).
\end{align}
\end{subequations}

%

\end{document}